\documentclass[fleqn,usenatbib]{mnras}
\usepackage{newtxtext,newtxmath}

\usepackage[T1]{fontenc}

\DeclareRobustCommand{\VAN}[3]{#2}
\let\VANthebibliography\thebibliography
\def\thebibliography{\DeclareRobustCommand{\VAN}[3]{##3}\VANthebibliography}

\usepackage{graphicx}	
\usepackage{amsmath}	

\usepackage{fix-cm}		
\usepackage{multirow}	
\usepackage{subfig}		

\usepackage{enumitem}

\newcommand\hi{\textsc{Hi}}

\title[Interaction-induced $\text{{\sc Hi}}$ gas concentration and central star formation]{Interaction-induced $\text{{\sc Hi}}$ gas concentration with centrally-enhanced star formation in ALFALFA-SDSS galaxies}

\author[Y. H. Guo \& C. Li]{
Yanhan Guo$^{1}$\thanks{E-mail: guoyh21@mails.tsinghua.edu.cn}
and Cheng Li$^{1}$\thanks{E-mail: cli2015@tsinghua.edu.cn}
\\
$^{1}$Department of Astronomy, Tsinghua University, Beijing 100084, China
}

\date{Accepted XXX. Received YYY; in original form ZZZ}

\pubyear{\the\year{}}

\begin{document}
\label{firstpage}
\pagerange{\pageref{firstpage}--\pageref{lastpage}}
\maketitle

\begin{abstract}
We present a statistical analysis for the interaction-induced central concentration of \hi\ gas distributions and its connection with interaction-induced central star formation enhancement, using a large sample of $\sim 10^4$ galaxies from the ALFALFA and SDSS surveys. By adopting the \hi\ profile parameter $K$, an indicator of gas concentration inferred from the integrated 21 cm emission line, we find that galaxies with more centrally concentrated \hi\ (higher $K$ values) or enhanced specific star foramtion rate (sSFR) exhibit significantly stronger clustering and higher probability of hosting a nearby neighbor on scales below $100h^{-1}\mathrm{kpc}$, which is more pronounced in low-mass galaxies. Furthermore, by utilizing the enhancement functions for a sample of galaxy pairs, we directly trace the evolution of \hi\ concentration and sSFR enhancement as a function of projected separation. Our findings indicate that tidal interactions drive a statistical synchrony between the central concentration of atomic gas and the enhancement of central star formation. Gas concentration appears to be a necessary condition for central star formation enhancement in interacting systems at all but the smallest separations. Compared to satellite galaxies, central galaxies exhibit stronger enhancement of gas fraction, gas concentration and sSFR, suggesting the role of environmental regulation.
\end{abstract}




\section{Introduction} 
\label{sec:intro}

Galaxy-galaxy interactions and mergers are fundamental drivers of galaxy formation and evolution. Numerical simulations have robustly demonstrated that tidal forces during such interactions can efficiently funnel gas  toward galactic centers, triggering central starbursts \citep[e.g.][]{Toomre1972,Barnes1991,Mihos1996, Springel2005,DiMatteo2007,Hopkins2008,Hopkins2013,Blumenthal2018,Moreno2021}). More recently,  cosmological simulations such as the Illustris-TNG project have extended these findings to more diverse galaxy populations, enabling systematic investigations across a wide range of masses, environments, and merger histories \citep[e.g.][]{Sparre2016, Patton2020,Faria2025}). While most studies have focused on total gas content, \citet{Faria2025} reported an enhancement in the gas fraction within the central stellar half-mass radius of galaxies, based on a large sample of mergering systems drawn from TNG cosmological simulations. 

Observationally, clear evidence for interaction-induced star formation enhancement in low-redshift galaxies has been established, thanks largely to large optical spectroscopic surveys, particularly the Sloan Digital Sky Survey (SDSS; \citealt{York2000}). For instance, \cite{Li2008a} applied multiple statistical methods---including two-point correlation functions, star formation rate enhancement functions and neighbor counts---to a sample of $\sim10^5$ star-forming galaxies from SDSS. They found that star formation rates in the central $1-2$ kpc of galaxies are enhanced when the separation between galaxies and their companions decreases.
Similarly, \citet{Ellison2008}) reported enhanced central star formation in galaxy-galaxy pairs compared to  control samples of isolated galaxies in SDSS. Numerous follow-up studies have examined the dependence of star formation enhancement on various galaxy and interaction properties, such as stellar mass, mass ratio, morphology, bulge mass, gas fraction, environment density, and spatial alignment \citep[e.g.][]{Ellison2010,Xu2010,Scudder2012,Scudder2015,Moon2019,He2022,Zee2024}. More recently, integral-field unit (IFU) surveys such as the Calar Alto Legacy Integral Field Area (CALIFA) survey and the SDSS-IV Mapping nearby Galaxies at Apache Point Observatory (MaNGA) survey have revealed that interaction-induced star formation can be spatially extended across entire galaxies \citep[e.g.][]{Barrera2015,Pan2019,Thorp2019,Jin2021,Steffen2021}. 

Observational studies of tidally induced cold gas inflows remain limited, however, largely due to the lack of suitable data. For individual galaxies, interferometric observations of CO emission have revealed centrally concentrated distribution of molecular gas (e.g. in NGC 1614 documented by \citealt{Konig2013, Saito2017}). By jointly analyzing IFU data from CALIFA and CO mapping from the EDGE survey for 58 nearby galaxies, \citet{Chown2019} found that the galaxies with centrally enhanced star formation are either barred or in mergers/pairs with relatively high concentrations of molecular gas. Their statistical results thus provide substantial evidence for the theoretical expectation that centrally enhanced star formation, previously identified  in optical  surveys, is fueled by the molecular gas transported inward from galactic disks via tidal interactions or bars. More recently, \citet{Thorp2022} investigated spatially resolved measurements of star formation rates and CO emission in 33 merging galaxies from the ALMaQUEST survey. They found that the centrally enhanced star formation can arise from diverse mechanisms, including an increased gas fraction due to molecular gas concentration and an enhanced star formation efficiency. Clearly, larger samples with both IFU and spatially resolved CO observations, covering a wide range of galaxy properties and merger stages, are needed to fully understand the relationships among interactions, mergers, star formation, and molecular gas distribution and inflow.

Compared to molecular gas, atomic gas is even more challenging to study spatially, owing to the coarse resolution of 21 cm emission line observations. To date, evidence for interaction-induced central concentration of \hi\ gas has been limited to either individual case studies (e.g., VCC479 by \citealt{Sun2025}) or small samples of interacting galaxies (e.g., \citealt{Holwerda2011}). 

In this paper, we aim to statistically investigate the interaction-driven central concentration of \hi\ gas in a large sample of galaxies from the Arecibo Legacy Fast ALFA survey \citep[ALFALFA;][]{ALFALFA2011, ALFALFA2018}, as well as its relationship with interaction-induced central star formation enhancement in the same galaxies. For this purpose, we employ the $K$ parameter measured by \citet{Yu2020, Yu2022b, Yu2022a} from the ALFALFA 21 cm emission line profiles to quantify the spatial concentration of \hi\ gas in galaxies. The $K$ parameter is defined as the integrated area enclosed by the normalized curve of growth and the diagonal line of unity. As demonstrated in previous work, $K$ serves as a reliable indicator of the \hi\ gas distribution: as $K$ increases, the 21 cm emission line profile transitions from a double-horned shape ($K<0$), characteristic of an extended \hi\ disk, to a single-peaked shape ($K>0$), indicative of a centrally concentrated \hi\ distribution. This parameter enables a statistical probe of \hi\ gas concentration in large galaxy samples for which spatially resolved \hi\ observations are unavailable. For instance, analyzing the $K$ parameter for 13,511 nearby galaxies from the SDSS and ALFALFA surveys, \citet{Yu2022a} found that galaxies with higher $K$ values—indicating more concentrated \hi\ gas—are associated with elevated total and central star formation. This finding strongly suggests a connection between \hi\ gas concentration and tidal interactions, given the well-established link between interactions and enhanced central star formation. The present work is motivated by this conjecture and aims to test it directly.

Following \citet{Li2008a}, we use SDSS fiber spectroscopy to probe star formation in the central $1-2$ kpc of galaxies. By applying multiple statistical methods—including projected two-point cross-correlation functions (2PCCFs), close neighbor counts, and enhancement functions—we quantify both the enhancement in star formation and the concentration of \hi\ gas in star-forming galaxies as a function of projected separation $r_p$ from their companions. The large ALFALFA–SDSS sample, combined with this multi-method approach, enables us to statistically link galaxy-galaxy interactions with central star formation activity and \hi\ gas concentration for the first time. As we demonstrate below, these analyses reveal a clear signature of centrally concentrated \hi\ gas during interactions, which is closely associated with interaction-induced star formation enhancement in the central region of galaxies.

This paper is organized as follows. In \autoref{sec:data}, we describe the sample selection from the ALFALFA survey and the construction of reference samples used to compute projected cross-correlation functions and close neighbor counts. In \autoref{sec:2ppcf}, we examine the dependence of galaxy clustering amplitude and neighbor counts on both the $K$ parameter and sSFR. In \autoref{sec:pair}, we construct a galaxy pair sample and directly compare the evolution of $K$ and sSFR during the interacting stage using the method of enhancement functions. In \autoref{sec:discussion}, we examine  the effects of unsettled gas and the central/satellite classification on our results, and discuss on implications and outlook of our work. Finally, we summarize our results in \autoref{sec:sum}. Throughout the paper, we adopt a $\Lambda$CDM cosmology with $\Omega_{m} = 0.3089$ and $h = 0.6774$, following \citet{Planck2016}.

\begin{figure*}
  \centering
  \begin{minipage}[t]{0.33\textwidth}
    \centering
    \includegraphics[width=\linewidth]{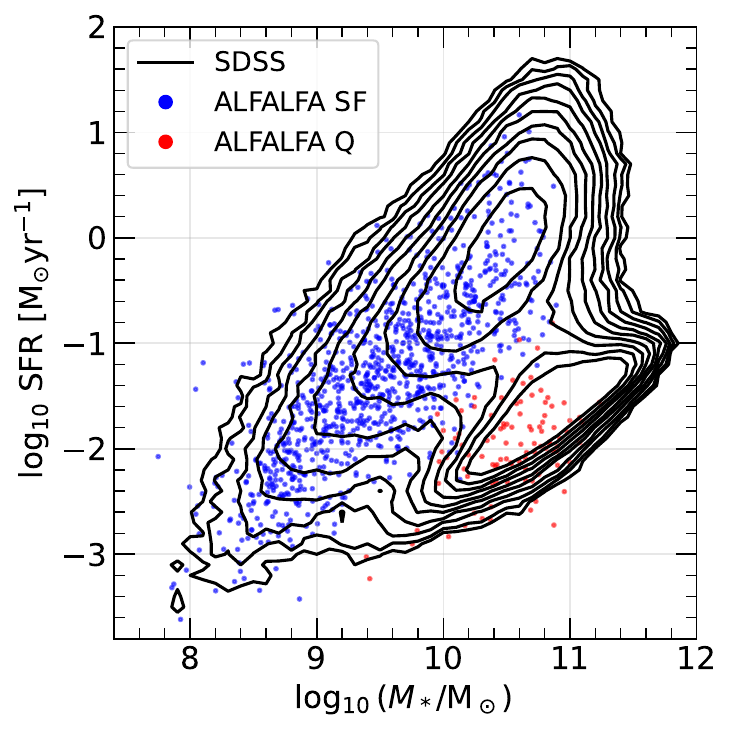}
  \end{minipage}
  \begin{minipage}[t]{0.4\textwidth}
    \centering
    \includegraphics[width=\linewidth]{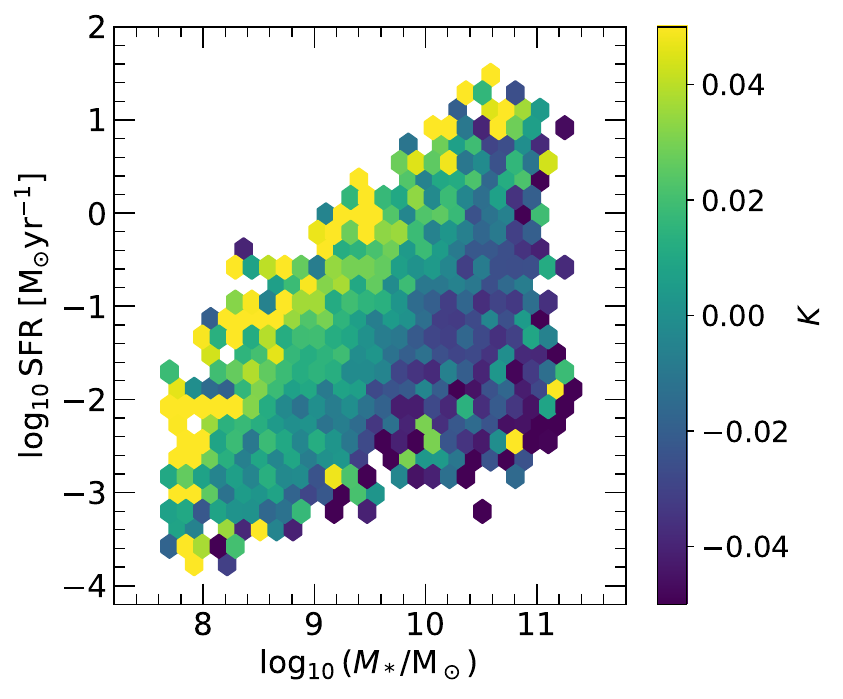}
  \end{minipage}
  \begin{minipage}[t]{0.33\textwidth}
    \centering
    \includegraphics[width=\linewidth]{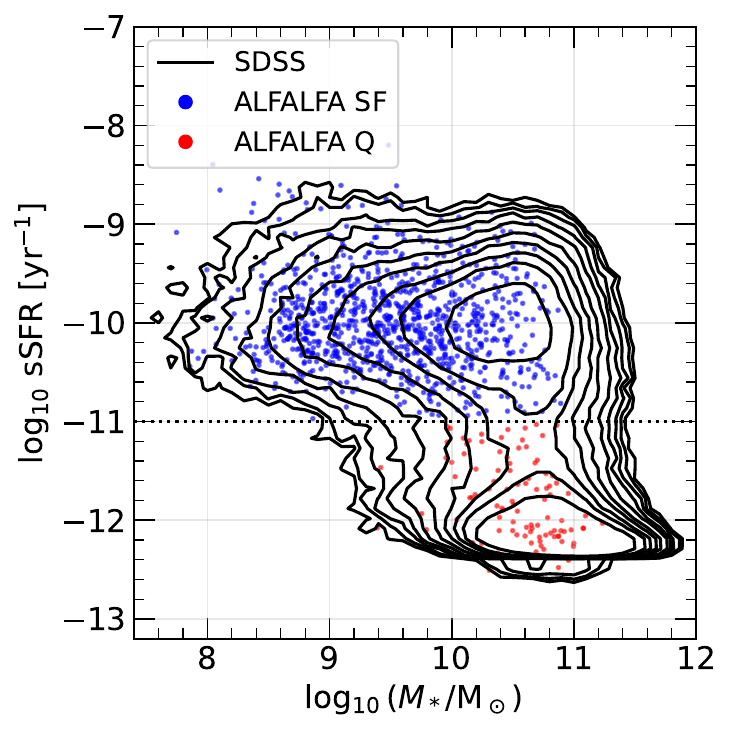}
  \end{minipage}
  \begin{minipage}[t]{0.4\textwidth}
    \centering
    \includegraphics[width=\linewidth]{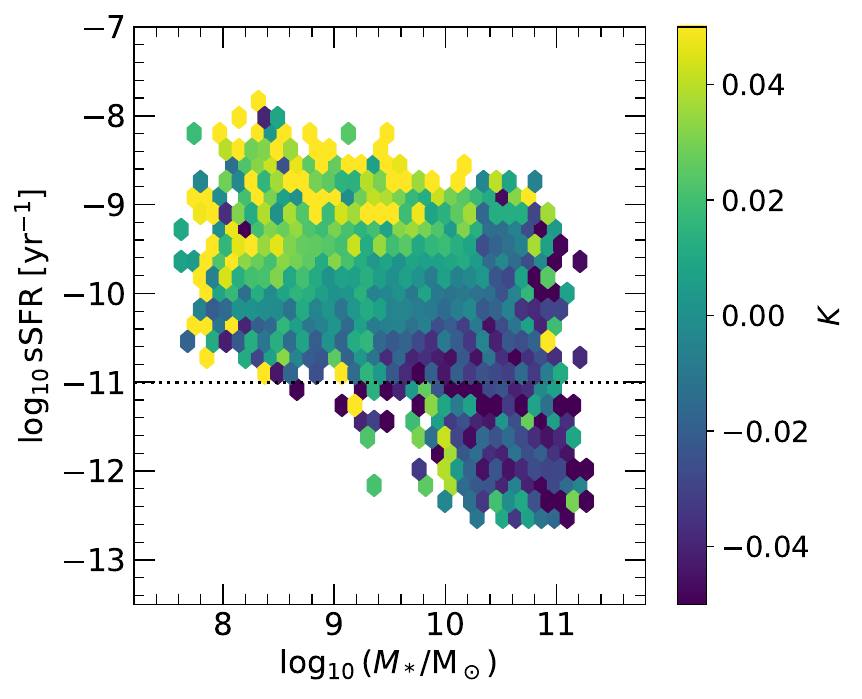}
  \end{minipage}
  \caption{The distribution of ALFALFA galaxies in the plane of stellar mass versus SFR (upper panel) and sSFR (lower panel). The left panels show the scatter plot, where galaxies are classified into star-forming (blue) and quenched (red) populations based on the dividing threshold of $\log_{10} \mathrm{sSFR} = -11.0 ~\mathrm{yr^{-1}}$. The black contours indicate the number density distribution of SDSS galaxies as the background. The corresponding right panels show the same distributions, with each pixel color coded by the median value of $K$.}
  \label{fig:contour_mass_sfrin}
\end{figure*}

\section{Samples} \label{sec:data}

\subsection{ALFALFA-SDSS star-forming galaxies}

The sample used in this paper is based on low-redshift ($z<0.06$) galaxies from the ALFALFA survey, after excluding objects without optical counterparts and those contaminated by radio frequency interference or nearby companions (\citealt{Yu2022b}). For each galaxy in this sample, we adopt the \hi\ profile parameter $K$ from \citet{Yu2022b} to quantify the gas concentration. To obtain optical counterparts, we cross-matched the ALFALFA sources with galaxies from the New York University Value-Added Galaxy Catalog (NYU-VAGC)\footnote{\url{http://sdss.physics.nyu.edu/vagc/}}, originally compiled by \citet{Blanton2005} based on the final data release of the SDSS \citep{SDSS-DR7}. Stellar masses are taken from NYU-VAGC, estimated by \citet{Blanton-Roweis-2007} from Petrosian magnitudes in the SDSS $u$, $g$, $r$, $i$, and $z$ bands. Star formation rates (SFRs) for the central regions of galaxies are obtained from the MPA-JHU catalog, which are derived from extinction-corrected H$\alpha$ emission and the $D_{n}4000$ index measured within the $3^{\prime\prime}$-diameter SDSS fiber \citep{Brinchmann2004}. Both stellar masses and SFRs assume the stellar initial mass function of \citet{Chabrier-2003}. Galaxies are classified into star-forming and quenched populations based on a specific star formation rate (sSFR) threshold of $\log_{10} \mathrm{sSFR} = -11.0~\mathrm{yr^{-1}}$. The resulting star-forming sample comprises 9594 galaxies and is used for the subsequent analysis.

In the left panels of \autoref{fig:contour_mass_sfrin}, we show the distribution of a randomly selected subset of 1000 galaxies from our sample in the stellar mass–SFR plane (upper panel) and the stellar mass–sSFR plane (lower panel). For comparison, the background contours display the number density distribution of galaxies from a reference sample selected from the SDSS spectroscopic survey (see below), which represents the general galaxy population in the local Universe. The right panels display the distribution of the $K$ parameter for our ALFALFA-SDSS star-forming sample in the same planes. As can be seen, at fixed stellar mass, galaxies with higher SFRs and sSFRs tend to exhibit higher $K$ values, consistent with the findings of \citet{Yu2022a}. This trend suggests that a more centrally concentrated distribution of \hi\ gas is closely associated with elevated central star formation activity in galaxies. 

\subsection{SDSS reference samples}
\label{sec:reference_samples}

Following \citet{Li2008a}, we construct two reference samples from NYU-VAGC. The first is a spectroscopic reference sample, comprising approximately half a million galaxies with $r$-band Petrosian apparent magnitudes $r<17.6$ and spectroscopic redshifts in the range $0.01<z<0.2$. The second is a photometric reference sample, consisting of roughly 3.5 million galaxies with $r$-band magnitudes in the range $10<r<19$. The spectroscopic reference sample is used to compute projected cross-correlation functions $w_{p}(r_{p})$ between the star-forming galaxies and reference galaxies, while the photometric reference sample is used to estimate neighbor counts around the star-forming galaxies. The measurements of $w_p(r_p)$ and neighbor counts will be presented in \autoref{sec:2ppcf}.

\subsection{Galaxy pairs and control samples}
\label{sec:pair_sample}

For each galaxy in our star-forming sample, we identify its closest companion in the NYU-VAGC, requiring a velocity difference $\Delta v < 500\,\mathrm{km\,s}^{-1}$ and a projected separation in the range $10\,h^{-1}\,\mathrm{kpc} < r_{p} < 250\,h^{-1}\,\mathrm{kpc}$. The lower bound on $r_{p}$ is adopted to exclude systems that are already in the merging stage, where the individual properties of the two galaxies can no longer be reliably measured. In addition, the Arecibo beam size of $3\farcm5$ in ALFALFA survey correspond to a separation of $\sim100h^{-1}\mathrm{kpc}$ at the median redshift $z=0.03$ of ALFALFA galaxies, which is comparable to the interaction scale of galaxy pairs. Therefore, we follow \cite{Yu2022b} to exclude galaxies that have a projected neighbor within the Arecibo beam whose \hi\ emission overlap with that of the target galaxy, resulting in an exclusion of $<10\%$ of the total galaxy pairs. For each paired galaxy, control galaxies are selected from the star-forming sample by extracting all galaxies that match criteria of 0.01 in redshift $z$, 0.1 dex in stellar mass $M_{\star}$, and 0.1 dex in local environmental density $\delta$. We evaluate the residuals $\Delta z$, $\Delta M_{\star}$, and $\Delta \delta$ as functions of projected pair separation and find that they remain approximately zero across all separations, demonstrating that the control sample successfully reproduces the distributions of $z$, $M_{\star}$, and $\delta$ of the paired galaxy sample. We restrict our pair sample to galaxies with at least 10 control galaxies, yielding a final sample of 2880 galaxy pairs and a median number of 53 control galaxies for the paired galaxies. In \autoref{sec:pair}, we will use this pair sample and the corresponding control samples to investigate the enhancement of sSFR and the $K$ parameter as functions of pair separation.

The environmental overdensity $\delta$ used above is taken from the ``Exploring the Local Universe with the reConstructed Initial Density Field'' project (ELUCID; \citealt{Wang2016}), which provides a high-resolution, constrained $N$-body simulation designed to reconstruct the density field of the nearby universe as traced by SDSS. In ELUCID, overdensity is defined as $\delta \equiv \rho/\bar{\rho} - 1$, where $\rho$ and $\bar{\rho}$ denote the local and mean matter densities, estimated by smoothing the density field with Gaussian kernels on scales of 2, 3, and $4\,h^{-1}\,\mathrm{Mpc}$. We adopt a smoothing scale of $2\,h^{-1}\,\mathrm{Mpc}$ to optimally capture the local environment. We have verified that using smoothing scales of 3 and $4\,h^{-1}\,\mathrm{Mpc}$ yields consistent results, indicating that our conclusions are not sensitive to the exact choice of smoothing scale.

\begin{figure*}
  \centering
  \includegraphics[width=0.45\linewidth]{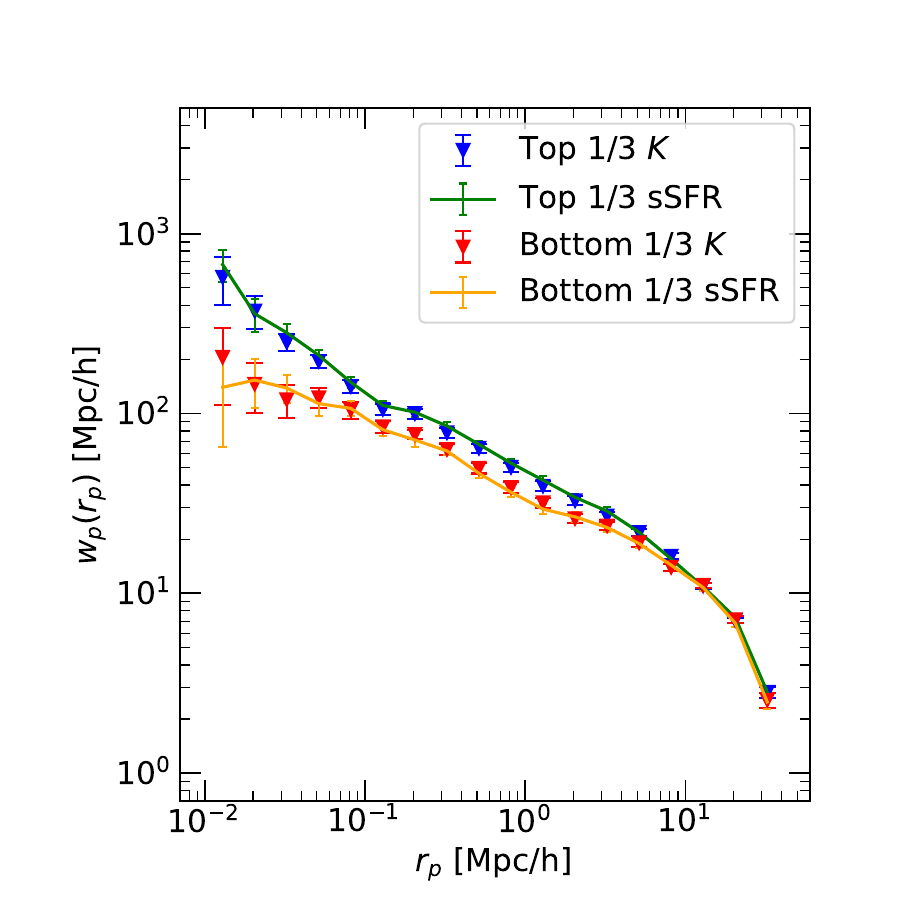}
  \includegraphics[width=0.45\linewidth]{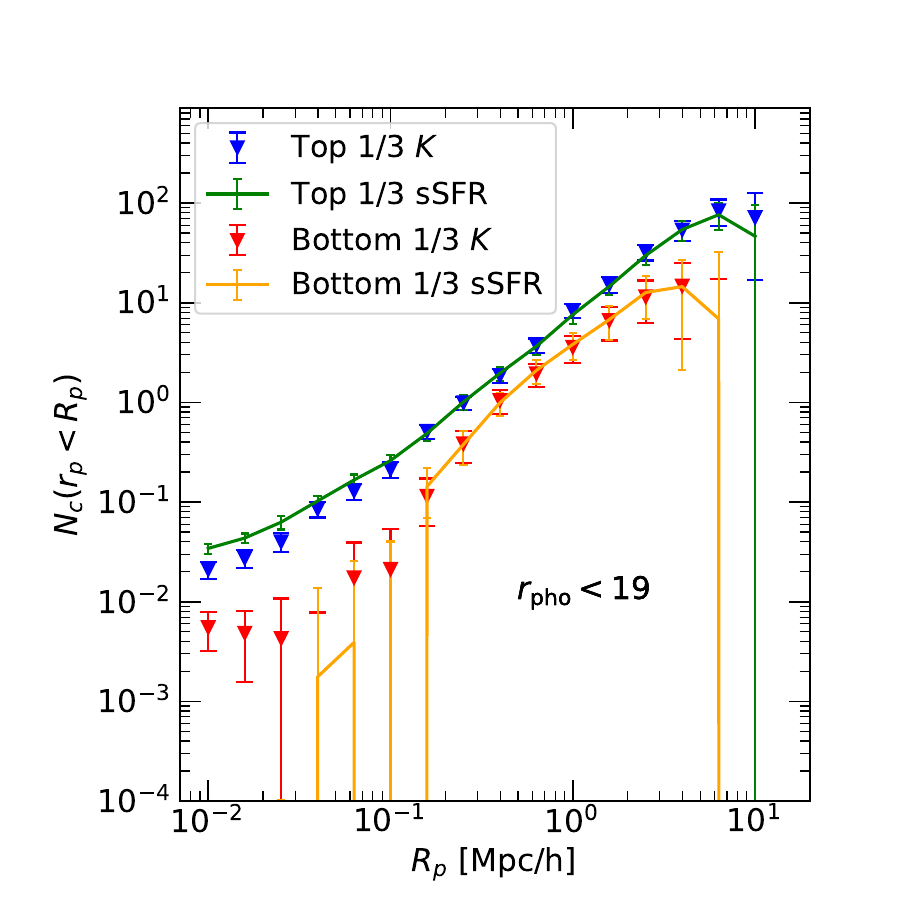}
  \caption{Projected cross-correlation functions (left panel) and close neighbor counts (right panel) for the two $K$-divided subsamples (shown by symbols), compared with the corresponding results for the $\mathrm{sSFR}$-divided subsample, as indicated.}
  \label{fig:2ppcf_sf_k_allmass}
\end{figure*}

\begin{figure*}
  \centering
  \includegraphics[width=1.0\textwidth]{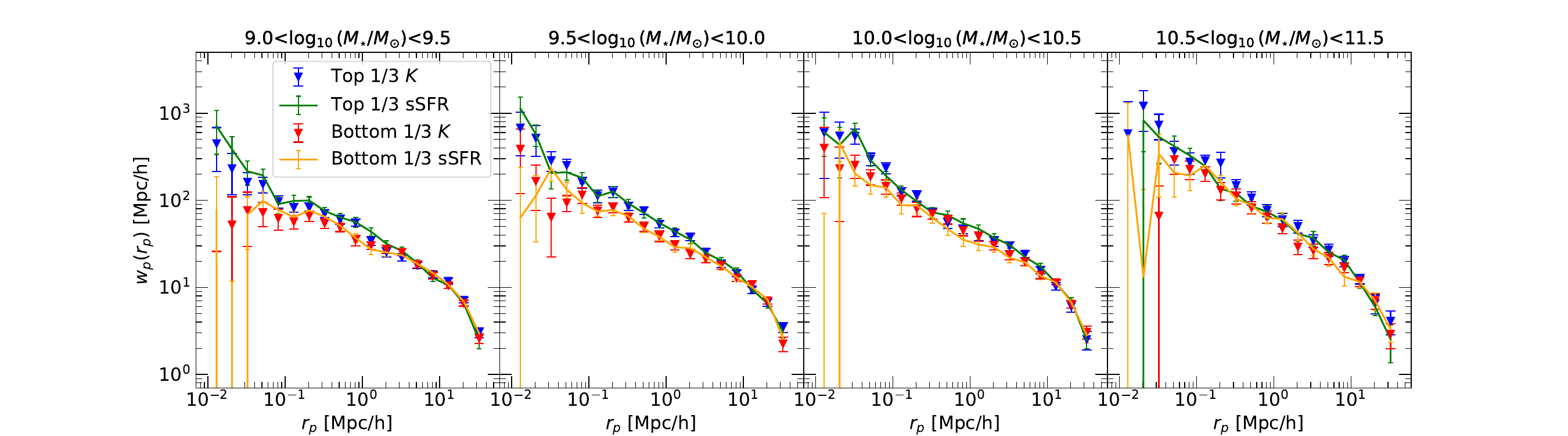}
  \includegraphics[width=1.0\textwidth]{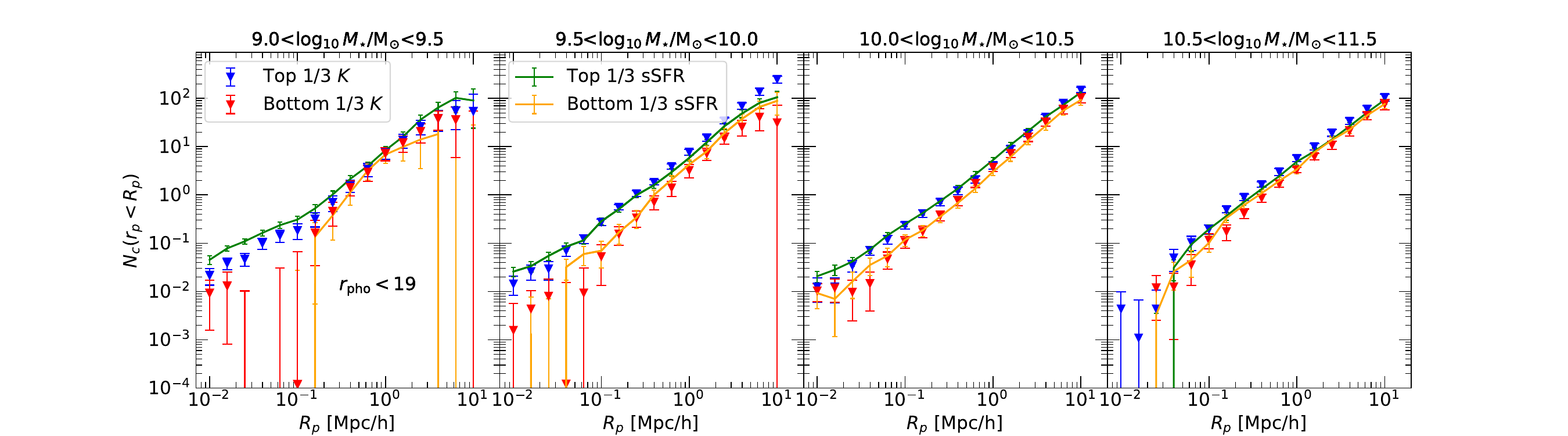}
  \caption{Same as \autoref{fig:2ppcf_sf_k_allmass}, for different mass intervals of star-forming sample, as indicated. }
  \label{fig:2ppcf_sf_k}
\end{figure*}

\section{Clustering and neighbor counts} \label{sec:2ppcf}

As demonstrated in previous studies \citep[e.g.][]{Li2006AGN,Li2008a,Li2008b}, the clustering amplitude on scales below $\sim 100\,h^{-1}\mathrm{kpc}$ can be used to probe galaxy-galaxy interactions. In this section, we employ both projected two-point cross-correlation functions (2PCCFs) and close  neighbor counts to examine the impact of interactions on central star formation activity—as traced by SDSS fiber spectroscopy—and on \hi\ gas concentration—as quantified by the \hi\ profile parameter $K$.

We adopt the methods of \citet{Li2006AGN} and \citet{Li2008a} to compute projected 2PCCFs and neighbor counts. Here we briefly outline the procedures and refer the reader to those papers for detailed descriptions and tests.

For a given (sub)sample of star-forming galaxies (hereafter {\tt Sample Q}), we first compute the redshift-space two-point cross-correlation function $\xi^{(s)}(r_{p},\pi)$ with respect to the spectroscopic reference sample ({\tt Sample D}; see \autoref{sec:reference_samples}) using the estimator
\begin{equation}
\xi^{(s)}(r_p,\pi) = \frac{N_{\text{R}}}{N_{\text{D}}}\cdot\frac{QD(r_p,\pi)}{QR(r_p,\pi)}-1,
\end{equation}
where $r_{p}$ and $\pi$ are the pair separations perpendicular and parallel to the line of sight, respectively; $N_{\text{D}}$ and $N_{\text{R}}$ are the numbers of galaxies in {\tt Sample D} and in a random sample ({\tt Sample R}) constructed to match the selection effects of {\tt Sample D}; and $QD(r_p,\pi)$ and $QR(r_p,\pi)$ are the cross-pair counts between {\tt Sample Q} and {\tt Sample D}, and between {\tt Sample Q} and {\tt Sample R}, respectively. The projected 2PCCF $w_p(r_p)$ is then obtained by integrating $\xi^{(s)}(r_p,\pi)$ along the line-of-sight direction over $\pi$ in the range $[-40,40]\,h^{-1}\mathrm{Mpc}$. We have corrected for the effect of fiber collisions following \citet{Li2006AGN}.

Neighbor counts are measured by counting galaxies in the photometric reference sample (see \autoref{sec:reference_samples}) around each star-forming galaxy within a given projected radius $R_{p}$, denoted as $N_{c}(r_{p}<R_{p})$. Background contamination is corrected by subtracting counts measured around randomly distributed points. To ensure a consistent lower bound on the luminosity contrast between neighbors and central galaxies across the full redshift range, we consider only neighbors brighter than $r_{\text{SFG}} + \Delta r$, where $\Delta r$ accounts for the difference in magnitude limits between the star-forming sample ($r<17.6$) and the photometric reference catalog ($r<19$). In our case, $\Delta r = 1.4$ mag.

We rank all galaxies in the star-forming sample in decreasing order of their $K$ values, and we construct two subsamples comprising the top third and the bottom third of the distribution. For each subsample, we then estimate the 2PCCF $w_p(r_p)$ and the neighbor counts $N_c$. In \autoref{fig:2ppcf_sf_k_allmass}, we present $w_{p}(r_{p})$ (left panel) and $N_{c}(r_{p}<R_{p})$ (right panel) for these two $K$-defined subsamples (shown as symbols with error bars). For comparison, we also rank galaxies by their central sSFR and compute both quantities for the top and bottom thirds of the sSFR distribution. The results for these sSFR-based subsamples are shown as green and yellow lines in both panels. All error bars are estimated using bootstrap resampling, based on 50 resamplings of each subsample of star-forming galaxies.

These measurements clearly show that galaxies with higher values of $K$ or sSFR tend to exhibit stronger clustering and higher neighbor counts on intermediate to small scales, with the effect being particularly pronounced at $r_p \lesssim 100\,h^{-1}\mathrm{kpc}$. In \autoref{fig:2ppcf_sf_k}, we further examine the dependence of $w_p(r_p)$ (upper panels) and $N_c(r_p<R_p)$ (lower panels) on both $K$ and sSFR, but now for different intervals of stellar mass (panels from left to right). The enhanced small-scale clustering and neighbor counts associated with higher sSFR and higher $K$, seen globally in the full sample, persist across all but the highest mass bins, though the effect becomes stronger at lower stellar masses.

\begin{figure}
  \centering
  \includegraphics[width=\linewidth]{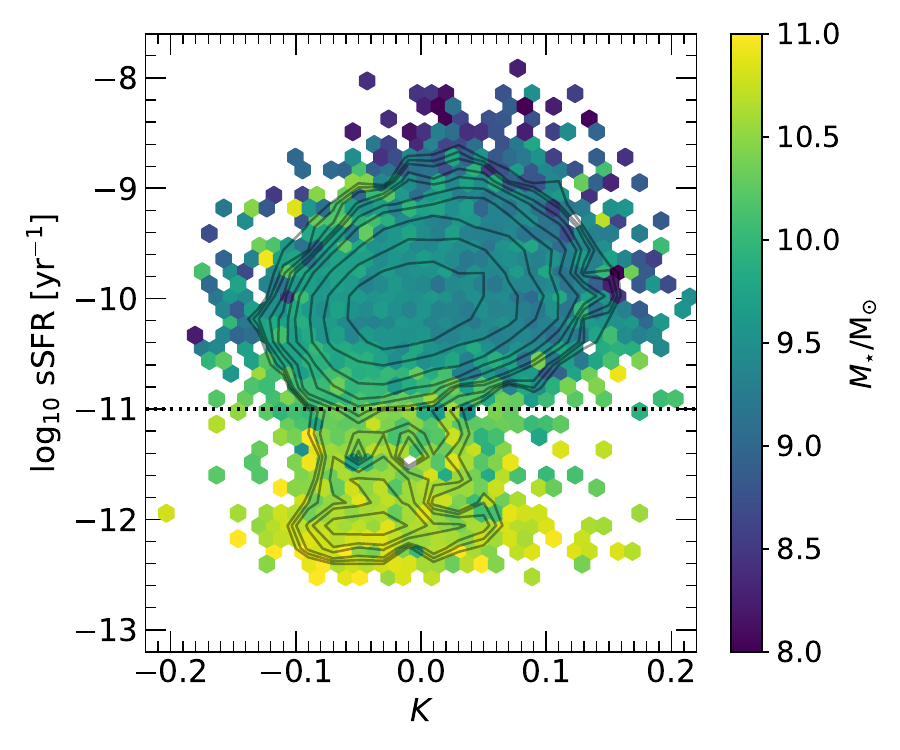}
  \caption{The distribution of ALFALFA galaxies in the $K$–sSFR plane with the contours representing the number density distribution of galaxies. Each pixel is color coded with the median value of stellar mass. The dotted line shows the sSFR threshold used to separate star-forming and quenched galaxies.}
 \label{fig:k_ssfrin_mass}
\end{figure}

The sSFR-related results are consistent with earlier findings by \citet{Li2008a}. What is new here are the results for subsamples selected by the $K$ parameter, which are remarkably similar to those defined by sSFR. Such close correspondence could, in principle, arise naturally if $K$ and sSFR were tightly correlated. To test this possibility, we show in \autoref{fig:k_ssfrin_mass} the distribution of all ALFALFA–SDSS galaxies in the $K$–sSFR plane. In the figure, the horizontal dotted line marks the sSFR threshold used to separate star-forming and quenched galaxies. Each pixel is color-coded by the median stellar mass of galaxies in that bin, illustrating the mass dependence, while the contours indicate the number density distribution. As can be seen, sSFR and $K$ exhibit only a weak correlation: galaxies with the highest sSFRs span a wide range in $K$, and vice versa. We have further examined the relation between $K$ and sSFR in different stellar mass intervals and find that $K$ shows a weak (though statistically significant) correlation with sSFR, with no strong dependence on stellar mass. This analysis demonstrates that the marked similarity in clustering and neighbor counts between the sSFR-defined and $K$-defined subsamples is not simply driven by a tight correlation between sSFR and $K$. The large scatter between sSFR and $K$ suggests that both parameters are influenced by a variety of processes, with tidal interactions being only one among many that may affect them. Therefore, to isolate the specific role of tidal interactions, it is more effective to directly examine interacting systems and their connection with $K$ and sSFR.

\begin{figure*}
  \centering
  \includegraphics[width=1.0\linewidth]{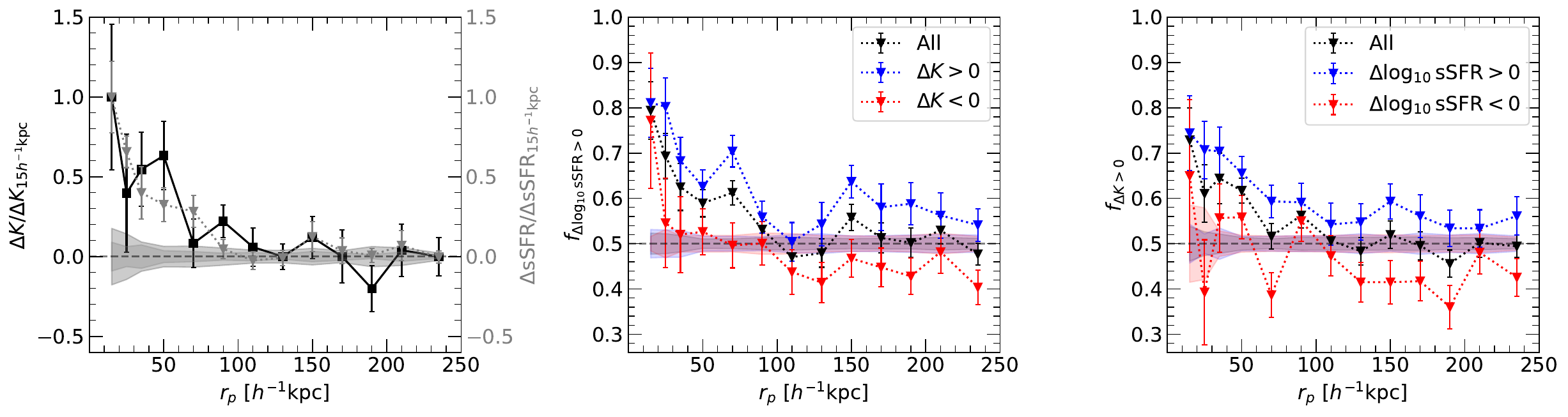}
  
  \caption{Left: Enhancement of $K$ (black solid line, normalized by the corresponding enhancement values of $r_p=15h^{-1}\text{kpc}$, referenced to the left y-axis) and sSFR (dotted gray line, referenced to the right y-axis) for the close galaxy pairs as a function of pair separations. Middle: The fraction of galaxies with enhanced sSFR as a function of pair separations, for the entire pair sample (black dotted lines), as well as for subsamples with gas concentration ($\Delta K>0$, blue dotted lines) and relatively diffuse gas ($\Delta K<0$, red dotted lines). Right: Similar as the middle panel, but for the fraction of galaxies with enhanced $K$ and the condition of $\Delta \log_{10}\mathrm{sSFR}$, as indicated. }
  \label{fig:Enhance_sSFR_rp_400kpc}
\end{figure*}

\section{Enhancements of sSFR and $K$ in paired galaxies} \label{sec:pair}

We quantify the effect of tidal interactions on a given galaxy property $Q$ by estimating the ``enhancement function'' $\Delta Q$ for each paired galaxy, using the pair galaxy sample and the corresponding control sample described in \autoref{sec:pair_sample}. The property $Q$ in our case is either sSFR or the $K$ parameter. Specifically, $\Delta Q(r_p)$ is defined as the relative difference in $Q$ between star-forming galaxies with pair separation of $r_p$ in the pair sample and their matched control galaxies:
\begin{equation}
\Delta Q(r_p) = Q_{\text{pair}}(r_p) - \tilde Q_{\text{ctrl}}(r_p)
\end{equation}
where $Q_{\text{pair}}(r_p)$ denotes the property $Q$ of the paired galaxy, and $\tilde{Q}_{\text{ctrl}}(r_p)$ is the median value of the corresponding control galaxies. Positive values of $\Delta Q$ indicate enhancement relative to the control sample, while negative values indicate suppression. This definition of $\Delta Q$ is similar to that used in previous studies \citep[e.g.][]{Ellison2008, Li2008a, Li2008b, Scudder2012}, although the  specific methods for defining control galaxies and/or computing  $\tilde{Q}_{\text{ctrl}}$ may vary across different studies.

The left panel of \autoref{fig:Enhance_sSFR_rp_400kpc} shows the median values of $\Delta K$ (referenced to the left y-axis; black solid line) and $\Delta \log_{10}\text{sSFR}$ (referenced to the right y-axis; gray dotted line) as functions of pair separation $r_p$. To enable a fair comparison between the two quantities, both $\Delta K$ and $\Delta \log_{10}\text{sSFR}$ at each $r_p$ are normalized by their respective values at the smallest separation bin. Error bars are estimated from 100 bootstrap resamples of the paired galaxies. To assess the uncertainty introduced by control sample selection, we perform an additional 100 bootstrap resamples of the control galaxies; the resulting $1\sigma$ uncertainties are shown as shaded regions—light gray for $K$ and dark gray for sSFR. As can be seen, both $K$ and sSFR exhibit increasing enhancement toward smaller pair separations within $\sim 100\,h^{-1}\,\mathrm{kpc}$. Notably, the enhancement of $K$ closely tracks that of sSFR, suggesting a synchronous evolution of gas concentration and star formation enhancement induced by galaxy-galaxy interactions.

The middle and right panels of the same figure display the fraction of galaxies with enhanced sSFR, $f_{\Delta \log_{10} \text{sSFR} > 0}$, and the fraction with concentrated gas, $f_{\Delta K > 0}$, as functions of $r_p$. In both panels, results are shown for the full paired sample (black dotted line), as well as for subsamples defined by the enhancement in the other parameter. Specifically, the middle panel shows results for subsamples with $\Delta K > 0$ (blue dotted line) and $\Delta K < 0$ (red dotted line), while the right panel shows subsamples with $\Delta \log_{10} \text{sSFR} > 0$ (blue dotted line) and $\Delta \log_{10} \text{sSFR} < 0$ (red dotted line). Error bars and shaded regions are derived from 100 bootstrap resamples of both paired galaxies and control galaxies.

Overall, both $f_{\Delta \log_{10} \text{sSFR} > 0}$ and $f_{\Delta K > 0}$ increase toward smaller pair separations, consistent with the enhancement functions shown in the left panel. Across all but the smallest separations, galaxies with $\Delta K > 0$ exhibit a significantly higher fraction of sSFR-enhanced galaxies, whereas for galaxies with $\Delta K < 0$, the fraction remains relatively flat and lies below the 50\% baseline. Similarly, across all but the smallest separations, galaxies with $\Delta \log_{10} \text{sSFR} > 0$ show a significantly higher fraction of gas-concentrated galaxies, while for those with $\Delta \log_{10} \text{sSFR} < 0$, the fraction remains flat and below 50\%. These behaviors strongly suggest that interaction-induced gas concentration and interaction-induced central star formation enhancement are closely linked: at all but the smallest pair separations, enhancement in either sSFR or $K$ appears to be a necessary condition for enhancement in the other.

\begin{figure*}
  \centering
  \includegraphics[width=1.0\linewidth]{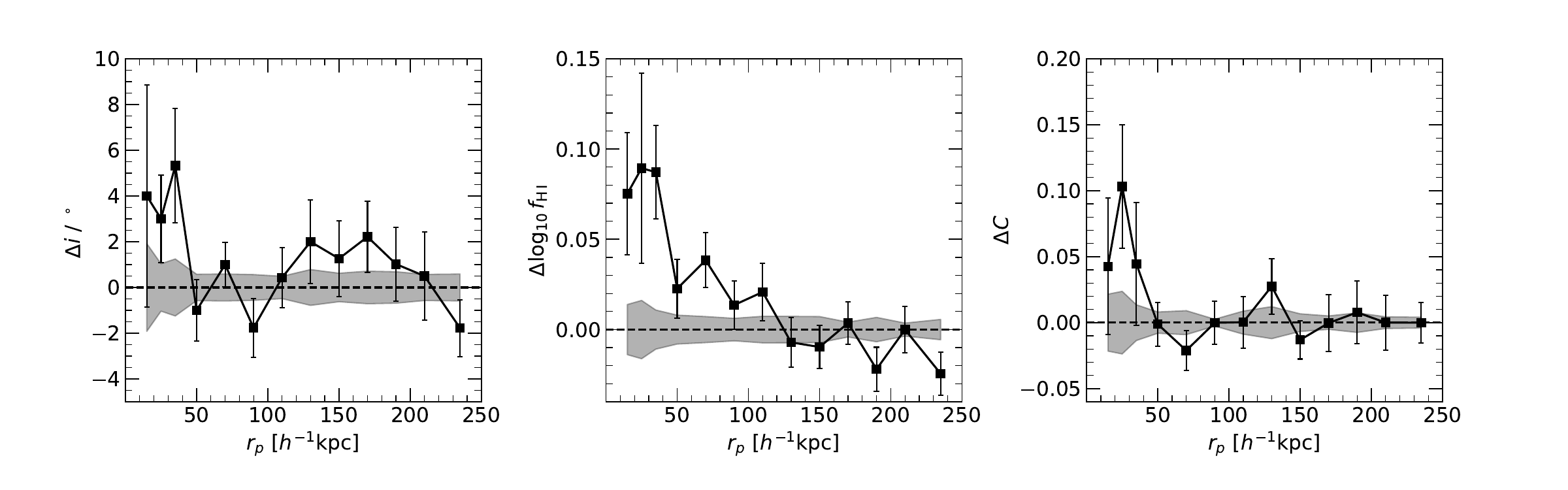}
  \caption{From left to right: enhancements of inclination angle $i$, gas fraction $f_{\mathrm{H\,I}}$ and optical concentration $C$, as a function of pair separations. }
  \label{fig:Enhance_hi_i_c}
\end{figure*}

At the smallest separation bin ($r_p \sim 15\,h^{-1}\,\mathrm{kpc}$), the middle panel shows that the fraction $f_{\Delta \log_{10} \text{sSFR} > 0}$ reaches $\sim 80\%$, independent of $\Delta K$, while the right panel shows that $f_{\Delta K > 0}$ reaches $\sim 65\%$, independent of $\Delta \log_{10} \text{sSFR}$. This result might suggest that, in strongly interacting systems where pair separations are comparable to galaxy sizes, there is no direct correlation between interaction-induced gas concentration and interaction-induced star formation enhancement. However, caution is warranted in drawing this conclusion. In such closely interacting systems, the cold gas distribution may be significantly disturbed by tidal forces, potentially leading to diverse outcomes including concentration, diffusion, or even gas loss. Moreover, the method for measuring the \hi\ profile parameter may become unreliable in these systems, where the kinematics of the \hi\ gas is substantially perturbed. In such cases, the $K$ parameter may no longer accurately reflect the true concentration of the \hi\ gas distribution. We will discuss on this issue in \autoref{sec:discussion}.

\begin{figure*}
  \centering
  \includegraphics[width=1.0\linewidth]{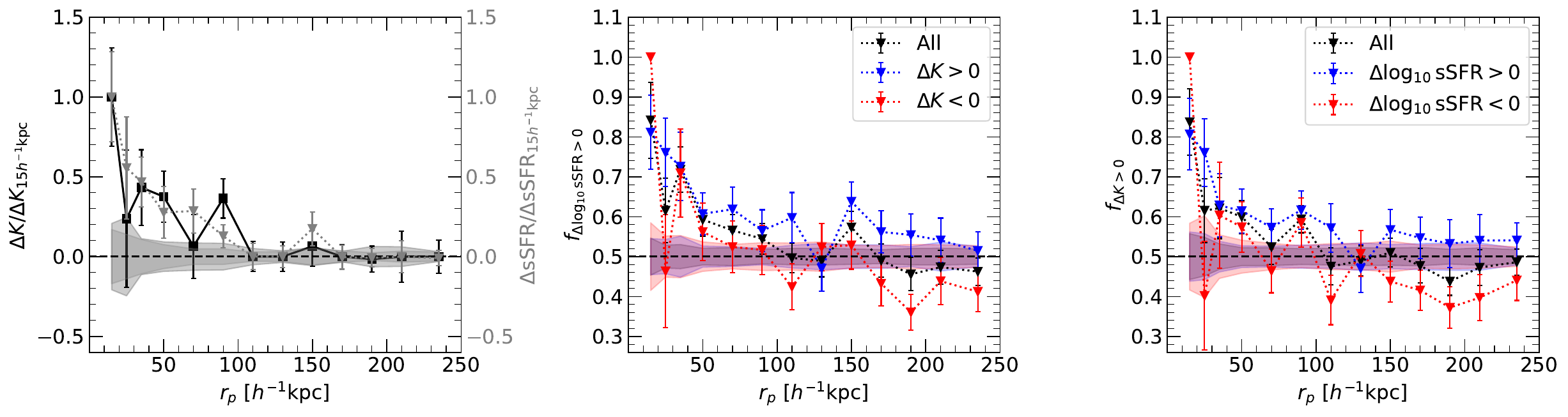}
  \caption{Same as \autoref{fig:Enhance_sSFR_rp_400kpc}, but with an addition control criterion applied to \hi\ mass.}
  \label{fig:Enhance_sSFR_rp_4d}
\end{figure*}

As noted by \citet{Yu2022a}, the integrated 21 cm emission line profile of a galaxy is not determined solely by its \hi\ radial distribution; it also depends on the size and inclination of the \hi\ disk, as well as the rotation curve. The \hi\ disk size can be estimated from the total \hi\ mass \citep[e.g.][]{Broeils1997, Wang2016HIsize}, and its inclination can be inferred from the axis ratio of the stellar light distribution under the assumption that gas and stars are coplanar. Estimating the rotation curve, however, is more challenging, as it depends on both the total mass and its distribution. In \autoref{fig:Enhance_hi_i_c}, we examine the differences between paired galaxies and their controls in three quantities: inclination angle $i$ (estimated from the $r$-band minor-to-major axis ratio, left panel), \hi\ gas fraction ($f_{\text{H\,{\sc i}}} \equiv M_{\mathrm{H,I}}/M_{\star}$, middle panel), and stellar light concentration ($R_{90}/R_{50}$, the ratio of radii enclosing 90\% and 50\% of the total $r$-band light, right panel). The gas fraction shows a significant enhancement within $\sim 100\,h^{-1}\,\mathrm{kpc}$, reaching about 0.1 dex at the smallest separations. In contrast, the enhancement in the inclination angle and stellar light concentration emerges only at very small separations ($\sim 40\,h^{-1}\,\mathrm{kpc}$) and remains modest, typically within 5$^{\circ} $ for inclination angle and 0.1 for stellar light concentration.

These results indicate that, compared to control galaxies matched in redshift, stellar mass, and environmental density, star-forming galaxies with close companions tend to have higher \hi\ gas fractions (implying larger \hi\ disks at fixed stellar mass) and more concentrated stellar light (suggesting a more steeply rising rotation curve in the central regions). An increase in stellar light concentration in paired galaxies was previously reported by \citet{Li2008a} and interpreted as a signature of structural changes driven by interaction-induced gas inflows and central star formation. However, this feature is relatively weak and emerges only at the smallest separations, so it is unlikely to play a major role in the enhancement of $K$ seen in earlier figures, which already becomes significant at much larger separations.

To isolate the effect of central gas concentration from that of the increase of global gas content, we therefore impose an additional control criterion on \hi\ mass $M_{\text{H\,{\sc i}}}$, requiring control galaxies to lie within 0.1 dex of each paired galaxy. After applying this criterion, we recompute the enhancement function for $f_{\text{H\,{\sc i}}}$ and find it consistent with zero at all pair separations, confirming that this tolerance effectively matches the gas fraction between paired and control galaxies. In \autoref{fig:Enhance_sSFR_rp_4d}, we compare the normalized enhancements of $K$ and sSFR (left panel) as well as the fractions of galaxies with enhanced sSFR ($f_{\Delta \log_{10} \text{sSFR} > 0}$, middle panel) and the fraction with concentrated gas ($f_{K>0}$, right panel), after applying this additional control on $M_{\text{H\,{\sc i}}}$. The new criterion does not significantly alter the enhancement functions and the fractions, indicating that the observed enhancement of $K$ reflects a genuine central concentration of \hi\ gas, rather than being driven by an increased global gas fraction. 

\begin{figure}
  \centering
  \includegraphics[width=0.8\linewidth]{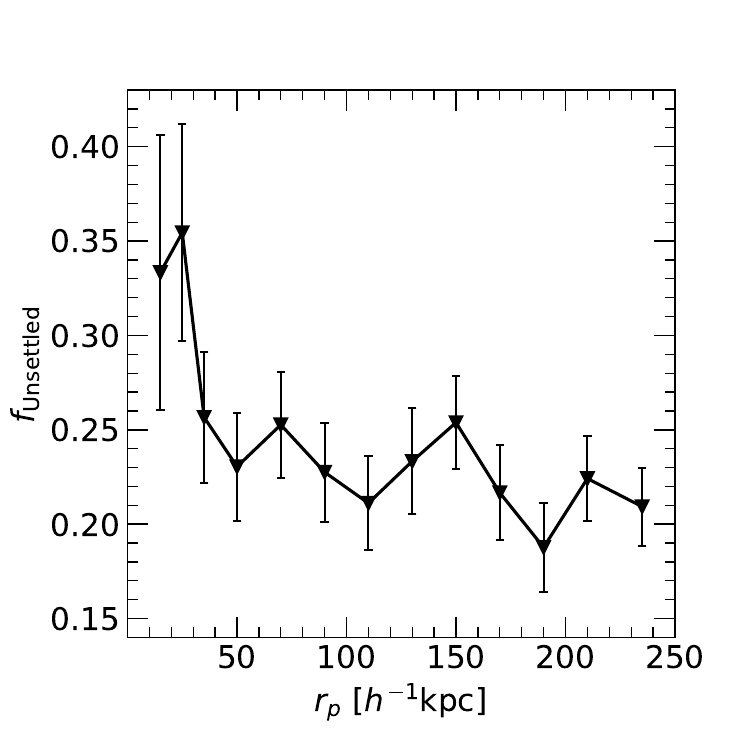}
  \caption{Fraction of paired galaxies with dynamically unsettled gas as a function of pair separations.}
  \label{fig:f_unsettled}
\end{figure}

\section{Discussion} \label{sec:discussion}

\subsection{Effect of unsettled gas}
\label{sec:unset}

\begin{figure*}
  \centering
  \includegraphics[width=1.0\linewidth]{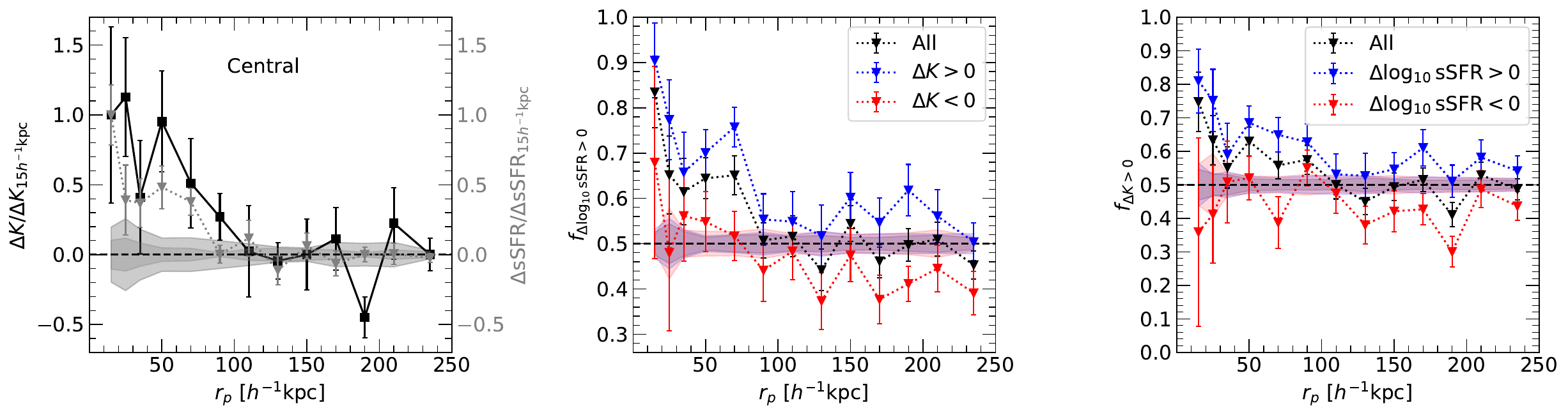}
  \includegraphics[width=1.0\linewidth]{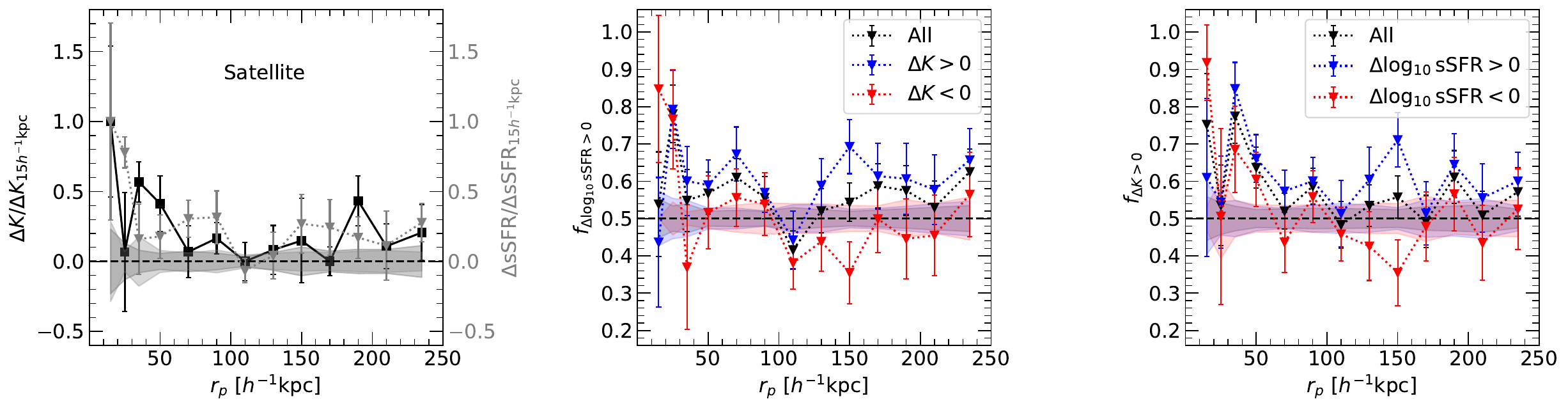}
  \caption{Same as \autoref{fig:Enhance_sSFR_rp_400kpc}, for central and satellite galaxies after excluding galaxies with dynamically unsettled gas, as indicated.}
  \label{fig:Enhance_cen_sat_sSFR_rp_400kpc}
\end{figure*}

As pointed out in \autoref{sec:pair}, central concentration is not the only possible outcome of the gas distribution in response to tidal interactions, particularly in strongly interacting systems. In such cases, an increase in the $K$ parameter may not be simply attributable to gas concentration. In a recent study, \citet{Huang2025} showed that an increase in $K$ can arise both from centrally concentrated \hi\ distributions and from dynamically unsettled \hi\ gas. They suggested that galaxies with unsettled \hi\ disks can be identified using the criterion $A_{F} > -2.9K + 1.3$, where $A_{F}$ is the flux asymmetry of the \hi\ profile \citep{Yu2022a}. \autoref{fig:f_unsettled} displays the fraction of galaxies in our pair sample that are classified as having unsettled gas, as a function of pair separation. At large separations ($\gtrsim 100\,h^{-1}\,\mathrm{kpc}$), the fraction remains relatively flat at $\sim 20\%$. Below this scale, the fraction increases with decreasing separation, with a particularly sharp rise at the smallest separations ($r_p < 40\,h^{-1}\,\mathrm{kpc}$). This trend is consistent with the expectation that strong tidal interactions can effectively disturb the \hi\ gas in galaxies.

We exclude galaxies with unsettled gas and repeat the entire analysis presented in the previous two sections. We find that our results remain largely unchanged, demonstrating that the uncertainty in $K$ contributed by galaxies with unsettled gas does not bias our conclusions.

\subsection{Effect of central/satellite classification}

Environmental effects occurring within dark matter halos can significantly reduce the gas content of satellite galaxies, which may also bias our results. To assess this, we examine whether the findings presented above depend on whether a galaxy is a central or satellite in its host dark matter halo. We obtain the central/satellite classification for galaxies in our sample from the SDSS galaxy group catalog constructed by \citet{Yang2007}. In \autoref{fig:Enhance_cen_sat_sSFR_rp_400kpc}, we present $\Delta \text{sSFR}$, $\Delta K$, and $f_{\Delta \log_{10} \text{sSFR} > 0}$ in the same manner as above—after applying the additional control on $M_{\text{H,I}}$ and excluding the paired galaxies with unsettled gas—but separately for central (upper panels) and satellite (lower panels) galaxies. As illustrated, while the synchronous evolution of gas concentration and sSFR enhancement persists for both populations, the enhancements in central galaxies exhibit a stronger increase toward small pair separations. In addition, both the fraction of sSFR-enhanced galaxies and its dependence on gas concentration are weaker for satellites. These results suggest that gas concentration and sSFR enhancements may be significantly modulated by the host halo environment, with satellites being more susceptible to environmental effects such as ram-pressure stripping. This is consistent with previous studies that have established that the cold gas fraction of satellite galaxies can be significantly reduced by environmental processes occurring within their host halos \citep[e.g.][]{Li2012, Zhang2013}.

\subsection{Influence of tidal interactions on gas fraction}

Gas fraction has been an active topic in studies of galaxy interactions. A number of works have reported enhanced molecular gas fractions in interacting galaxies (e.g., \citealt{Saintonge2012, Violino2018, Pan2018, Yu2024}). In contrast, the behavior of \hi\ gas remains a subject of debate. Some studies have found enhanced \hi\ gas fractions in interacting systems \citep{Jaskot2015, Ellison2018}, while others report \hi\ content comparable to that of isolated galaxies (e.g., \citealt{Zuo2018, Bok2020, Jin2025}) or even marginally reduced values \citep{Yu2022}. These diverse findings suggest that the \hi\ content during galaxy interactions is governed by multiple competing processes, including consumption by star formation, stripping by tidal force, as well as gas accretion through cooling of the surrounding intergalactic medium \citep[e.g.][]{Wang2023}. 

In our work, we consistently detect an overall increase in \hi\ gas fraction in paired galaxies (see \autoref{fig:Enhance_hi_i_c}). Previous studies have shown that the \hi\ gas content can also be influenced by environmental effects, with central galaxies preferentially exhibiting elevated \hi\ content \citep{Janowiecki2017,Yan2026} and satellite galaxies being \hi\ deficient (e.g., \citealt{Verdes2001,Li2012,Zhang2013}).
Consistent with previous findings, the increase of gas fraction in galaxy pairs becomes weaker when the analysis is restricted to satellite galaxies. Thus, whether the gas fraction increases in interacting galaxies depends on the balance among these competing mechanisms. Importantly, we find that the enhancement of central star formation in close pairs is not driven by an increased global gas fraction, but rather by interaction-induced gas concentration. We note, however, that our sample is biased toward relatively \hi\,-rich galaxies, as it is selected from the ALFALFA survey; our conclusions therefore apply primarily to such galaxies. Extending this analysis to galaxies with lower \hi\ gas content would be a valuable direction for future work.

\subsection{A synchronous evolution driven by interactions}

The highly similar trends of $\Delta K$ and $\Delta \log_{10}\text{sSFR}$ with pair separation found in our work strongly suggest a synchronous evolution of gas concentration and central star formation enhancement, both induced by galaxy-galaxy interactions. Most studies in the literature have focused on establishing evidence for interaction-induced gas inflows, enhanced star formation, and their correlations, with relatively few examining the temporal sequence between these two processes. In numerical simulations, star formation is typically triggered once the gas density exceeds a threshold, leading to comparable evolutionary timescales for gas concentration and star formation enhancement. For instance, from the figures presented in \citet{Faria2025}, which analyzed a large merger sample drawn from the TNG cosmological simulations, one can see that the central sSFR increases on timescales similar to those of the central gas fraction. Our results provide observational support for this picture.

As mentioned in \autoref{sec:intro}, this work was motivated by \citet{Yu2022a}, who found an intriguing association between higher $K$ values—indicative of more concentrated \hi\ gas—and elevated star formation levels in nearby galaxies. However, the physical origin of the \hi\ gas concentration itself remained an open question. Our work now provides a firm answer to that question, demonstrating that galaxy-galaxy interactions are a key driver of both \hi\ gas concentration and the accompanying enhancement in central star formation.

\subsection{Outlook}

While single-dish \hi\ observations such as ALFALFA used in our work have provided important statistical constraints, their limited spatial resolution prevents more direct measurements of the internal gas distribution within galaxies. Interferometric \hi\ surveys with current and upcoming facilities, such as VLA, ASKAP, MeerKAT, and SKA, will enable studies of spatially resolved atomic gas in large samples of interacting galaxies. These observations are crucial for directly tracing interaction-induced gas inflows and constraining the physical link between gas concentration and star formation.

\section{Summary} \label{sec:sum}

In this paper, we investigate the response of the \hi\ gas distribution to galaxy-galaxy interactions in nearby star-forming galaxies, using a large sample of $\sim 10^4$ galaxies from the ALFALFA and SDSS surveys. To characterize the central concentration of \hi\ gas, we employ the \hi\ profile parameter $K$ provided by \citet{Yu2022b}. We first compute projected two-point cross-correlation functions $w_p(r_p)$ and close neighbor counts $N_c(r_p<R_p)$ for star-forming galaxies with different specific star formation rates (sSFRs, measured from SDSS fiber spectroscopy and thus tracing the central $1-2$ kpc) and different $K$ values, to examine how clustering and neighbor counts depend on these properties. We then construct a sample of star-forming galaxies each with a close companion and estimate the enhancements of sSFR and $K$ by comparing paired galaxies with control samples closely matched in redshift, stellar mass, environmental overdensity, and total \hi\ gas mass. These complementary approaches enable us to quantify the effects of tidal interactions on gas concentration and star formation enhancement as a function of decreasing pair separation. We also examine the potential effect of the subset of galaxies with unsettled gas and the central/satellite classification on our results. 

The main results of this paper can be summarized as follows:

\begin{enumerate}[label=(\roman*)]
\item Star-forming galaxies with higher values of $K$ (indicative of a more centrally concentrated distribution of \hi\ gas) or sSFR exhibit stronger clustering and a higher probability of hosting a nearby neighbor on scales below $\sim 100\,h^{-1}\,\mathrm{kpc}$. This effect is more pronounced for lower-mass galaxies. The marked similarity in  the dependence of clustering and neighbor counts on sSFR and $K$ is not simply driven by a tight correlation between sSFR and $K$.

\item Both central sSFR and gas concentration, as traced by $K$, are significantly enhanced in paired star-forming galaxies with pair separations smaller than $\sim 100\,h^{-1}\,\mathrm{kpc}$, relative to control galaxies matched in redshift, stellar mass, and environmental overdensity. 

\item The enhancements of gas concentration and sSFR evolve synchronously as pair separation decreases. This result persists even when control galaxies are additionally matched in \hi\ gas mass, indicating that the enhancement of $K$ reflects a genuine central concentration of \hi\ gas, rather than being driven by an increase in global gas fraction.

\item Galaxies with $\Delta K>0$ exhibit a significantly higher fraction of galaxies with enhanced sSFR, while for those with $\Delta K<0$, the fraction remains relatively flat and lies even below the average. Similarly, galaxies with $\Delta \log_{10}\text{sSFR}>0$ exhibit a higher fraction of galaxies with concentrated gas, while for those with $\Delta \log_{10}\text{sSFR}<0$, the fraction remains flat and low. This implies that enhancement in either sSFR or $K$ appears to be a necessary condition for enhancement in the other.

\item Our conclusions remain largely unchanged after excluding galaxies with unsettled gas or dividing the galaxies according to their central/satellite classification. The enhancements of both gas concentration and sSFR are more pronounced for central galaxies than for satellite galaxies, highlighting the role of environmental effects within dark matter halos—such as ram-pressure stripping—that can reduce the cold gas fraction and weaken the concentration of gas distribution.
\end{enumerate}

\section*{Acknowledgements}
This work is supported by the National Key R\&D Program of China (grant NO. 2022YFA1602902), the National Natural Science Foundation of China (grant Nos. 12433003), and China Manned Space Program with grant no. CMS-CSST-2025-A10.

Funding for SDSS and SDSS-II has been provided by the
Alfred P. Sloan Foundation, the Participating Institutions, the
National Science Foundation, the U.S. Department of Energy,
the National Aeronautics and Space Administration, the
Japanese Monbukagakusho, the Max Planck Society, and the
Higher Education Funding Council for England. The SDSS
website is (\url{http://www.sdss.org/}). SDSS is managed by the Astrophysical Research Consortium for the Participating Institutions. The Participating
Institutions are the American Museum of Natural History,
Astrophysical Institute Potsdam, University of Basel, University of Cambridge, Case Western Reserve University,
University of Chicago, Drexel University, Fermilab, the
Institute for Advanced Study, the Japan Participation Group,
Johns Hopkins University, the Joint Institute for Nuclear
Astrophysics, the Kavli Institute for Particle Astrophysics and
Cosmology, the Korean Scientist Group, the Chinese Academy
of Sciences (LAMOST), Los Alamos National Laboratory, the
Max-Planck-Institute for Astronomy (MPIA), the Max-PlanckInstitute for Astrophysics (MPA), New Mexico State University, Ohio State University, University of Pittsburgh,
University of Portsmouth, Princeton University, the United
States Naval Observatory, and the University of Washington.

We acknowledge the Tsinghua Astrophysics High-Performance Computing platform at Tsinghua University for providing computational and data storage resources that have contributed to the research results reported within this paper.

\section*{Data Availability}

The data used in this work include the following: (1) the SDSS-based galaxy catalogs, specifically the New York University Value-Added Galaxy Catalog (NYU-VAGC), compiled by \citet{Blanton2005} and available at \url{http://sdss.physics.nyu.edu/vagc/}; (2) the MPA-JHU SDSS catalog, described in \citet{Brinchmann2004} and accessible at \url{https://wwwmpa.mpa-garching.mpg.de/SDSS/}; (3) the \hi\ profile parameters measured by \citet{Yu2022b} (see their Table 2 and Table 3) based on the ALFALFA survey data; and (4) the environmental overdensity $\delta$ for SDSS galaxies, provided by the ``Exploring the Local Universe with the reConstructed Initial Density Field'' project (ELUCID; \citealt{Wang2016}).



\bibliographystyle{mnras}
\bibliography{example} 






\bsp	
\label{lastpage}
\end{document}